\documentclass[aps,prb,twocolumn]{revtex4} 

\usepackage{graphicx}
\usepackage{dcolumn}
\usepackage{bm}
\usepackage{amsmath} 

\newcommand{\comment}[1]{}

\begin{document}
\renewcommand{\theequation}{\arabic{section}.\arabic{equation}}

\title{Quantum Statistical Mechanics in Classical Phase Space. V.
Quantum Local, Average Global}


\author{Phil Attard}
\affiliation{{\tt phil.attard1@gmail.com}}


\begin{abstract}
One-particle energy eigenfunctions
are used to obtain quantum averages in many particle systems.
These are based on the effective local field due to fixed neighbors
in classical phase space,
while the averages account for the non-commutativity
of the position and momentum operators.
Used in Monte Carlo simulations for a one-dimensional Lennard-Jones fluid,
the results prove more reliable
than a high temperature expansion
and a harmonic local field approach,
and at intermediate temperatures agree with benchmark numerical results.
Results are presented for distinguishable particles, fermions, and bosons.
\end{abstract}

\pacs{}

\maketitle

%
\section{Introduction}
\setcounter{equation}{0} \setcounter{subsubsection}{0}
%

The problem with doing quantum mechanics for a many particle system
is that it is prohibitively expensive to compute
the energy eigenfunctions for the entire system.
One should avoid this if at all possible.
This is one reason for formulating quantum statistical mechanics
in classical phase space.

Another reason lies in the observation
that terrestrial condensed matter is primarily classical,
with quantum corrections being relatively small in typical examples
(eg.\ ${\cal O}(10^{-4})$ in water at standard temperature and pressure).
Of course there are systems where quantum effects are larger,
and even cases where they are dominant,
but it makes more sense to treat many-particle systems
as classical systems with successive quantum corrections
than it does to treat them as fully quantum mechanical systems
in which classical phenomena emerge after some sort
of infinite re-summation or averaging process.
It would be better to compute quantum effects in some local region
and to extend these to the whole system,
than to require the wave function for the entire system.

In the formally exact transformation
of quantum statistical mechanics to classical phase space
two phase functions arise:
\cite{STD2,Attard18a}
the commutation function,
which accounts for the non-commutativity
of the position and momentum operators,
and the symmetrization function,
which accounts for the wave function symmetry
of bosons and fermions.
The symmetrization function is relatively trivial to compute,
since it amounts to little more than a Fourier exponential factor
for permutations amongst neighboring particles.
Accordingly, recent work by the author
has focussed on the commutation function
with a view to obtaining
practical, efficient, and accurate computational formulations of it.

The present author's commutation function\cite{STD2,Attard18a}
is in essence the same as a function introduced by Wigner;\cite{Wigner32}
the difference lies in the formulation of the wave functions
and the consequent weight for classical phase space.
Wigner\cite{Wigner32} and Kirkwood\cite{Kirkwood33}
used this function
to obtain the primary quantum correction to classical thermodynamics,
which is the first term in a high temperature series expansion.
The present author extended this series to fourth order
and applied to a Lennard-Jones fluid in one- and in three-dimensions.
\cite{Attard16,Attard18b,Attard19b}
Based on these results it was concluded that
the high temperature expansion is too unwieldy
and too slowly converging under terrestrial conditions
to be a practical approach in the long term
(see also the results below).

Moving forward,
the author next explored a so-called mean field approximation,
based on a second order expansion of the potential energy
about a local minimum for the current configuration,
together with the exact commutation function
of a simple harmonic oscillator.\cite{Attard18b}
(Below this is called the harmonic local field approach,
which is more precise than `mean field'.)
This  proved quite accurate for a one-dimensional harmonic crystal
at all temperatures,\cite{Attard19b}
and also for a one-dimensional Lennard-Jones fluid at high
temperatures.\cite{Attard18c}
But, as is shown here,
it is less successful for the latter at
intermediate and low temperatures.

In more recent theoretical work,\cite{Attard19c}
the author has proposed that the commutation function
could be written as a series of one-body, two-body,  etc.\
effective potentials.
In implementing that approach numerical problems were encountered.
Nevertheless the  one-body idea
has led to a modified approach for the present paper.
The commutation function is here written
as the sum over energy eigenfunctions for one variable particle
in the effective local field from  fixed neighbor particles.
This is a little like a Born-Oppenheimer approach,
except that the effective rather than the actual local field is used,
and the energy eigenfunctions of the total system are not sought.
However the singlet energy eigenfunctions
that are here obtained give a realistic singlet commutation function
whose sum gives the  full system  phase space weight.
Although the commutation function is referred to as singlet,
the presence of the fixed neighbors
really make it a many-body effective potential.

The final algorithm turns out to be almost identical
to the earlier harmonic local field (then called mean field) approach,
with one improvement.
The singlet commutation function
is now calculated numerically for the exact effective local field.
The earlier approach used a harmonic approximation
to the effective local field.
What the two approaches have in common
is that the effective local field
contains half the pair contribution
compared to the actual local field
(see \S~\ref{Sec:LocalField}). 


%
\section{Algorithm}
\setcounter{equation}{0} \setcounter{subsubsection}{0}
%

\subsection{Phase Space Formulation of Quantum Statistical Mechanics}

The  quantum probability density in classical phase space
for the canonical equilibrium system is\cite{STD2,Attard18a}
\begin{equation}
\wp^\pm({\bf q},{\bf p})
=
\frac{ e^{-\beta {\cal H}({\bf q},{\bf p})}}{N! h^{3N} Z^\pm(T)}
\, \omega({\bf q},{\bf p}) \, \eta^\pm({\bf p},{\bf q}).
\end{equation}
Here $N$ is the number of particles,
${\bf p} = \{{\bf p}_1,{\bf p}_2,\ldots,{\bf p}_N\}$
are their momenta,
with ${\bf p}_j = \{p_{jx},p_{jy},p_{jz}\}$,
 ${\bf q} = \{{\bf q}_1,{\bf q}_2,\ldots,{\bf q}_N\}$
are the positions,
$\beta = 1/k_\mathrm{B}T$ is the inverse temperature,
with $T$ the temperature
and $k_\mathrm{B}$ Boltzmann's constant,
and $h$ is Planck's constant.
The classical Hamiltonian is
${\cal H}({\bf q},{\bf p}) = {\cal K}({\bf p}) + U({\bf q})$,
with ${\cal K} = p^2/2m$ being the kinetic energy,
and $U$ being the potential energy.
The  partition function $Z^\pm(T)$ normalizes
the phase space probability density.
The quantum aspects are embodied in the symmetrization function $\eta^\pm$
and the commutation function density $\omega$.

The symmetrization and commutation functions
are defined in terms
of the unsymmetrized position and momentum eigenfunctions,
which in the position representation ${\bf r}$
are respectively\cite{Messiah61}
\begin{equation} \label{Eq:qrpr}
|{\bf q}\rangle = \delta({\bf r}-{\bf q})
, \mbox{ and }
|{\bf p}\rangle
=
 e^{-{\bf p}\cdot{\bf r}/i\hbar} , 
\end{equation}
where $\hbar=h/2\pi$.
The normalization of the momentum eigenfunctions is
unimportant as only their ratio appears below.

The symmetrization function is formally\cite{Attard18a,Attard19a}
\begin{equation}
\eta^\pm({\bf p},{\bf q})
\equiv
\frac{1}{\langle {\bf p} | {\bf q} \rangle }
\sum_{\hat{\mathrm P}} (\pm 1)^p \,
\langle \hat{\mathrm P} {\bf p} | {\bf q} \rangle ,
\end{equation}
with $\hat{\mathrm P}$  the permutation operator
and $p$ its parity.
The plus sign is for bosons and the minus sign is for fermions.


The commutation function density $\omega$,
which is essentially the same as the functions
introduced by Wigner\cite{Wigner32}
and analyzed by Kirkwood,\cite{Kirkwood33}
is defined by
\begin{equation} \label{Eq:def1-Wp}
e^{-\beta{\cal H}({\bf q},{\bf p})}
\omega({\bf q},{\bf p})
=
\frac{\langle{\bf q}|e^{-\beta\hat{\cal H}} |{\bf p}\rangle
}{\langle{\bf q}| {\bf p}\rangle }.
\end{equation}
Using the completeness of the energy eigenfunctions,
$\hat{\cal H} | {\bf n}\rangle = E_{\bf n} | {\bf n}\rangle $,
this can be rewritten as
\begin{equation}\label{Eq:def-Wp}
e^{-\beta{\cal H}({\bf q},{\bf p})}
\omega({\bf q},{\bf p})
=
\frac{1}{\langle{\bf q}| {\bf p}\rangle }
\sum_{\bf n}
e^{-\beta E_{\bf n}}
\langle{\bf q}| {\bf n}\rangle\,
\langle{\bf n}| {\bf p}\rangle .
\end{equation}
The departure from unity of the commutation function
reflects the non-commutativity of the position and momentum operators.
Obviously the system must become classical in the high temperature limit,
and so $\omega({\bf q},{\bf p}) \rightarrow 1$, $\beta \rightarrow 0$.


The commutation function is usefully cast as
a temperature-dependent effective potential,
\cite{STD2}
\begin{equation} \label{Eq:expw}
\omega({\bf q},{\bf p}) \equiv e^{W({\bf q},{\bf p})} .
\end{equation}
(This notation differs from previous articles.)
The rationale for this is
that $W$ is extensive with system size,
which is an exceedingly useful property in thermodynamics.

\subsection{Effective Local Field Formulation for the Commutation Function}
\label{Sec:LocalField}

In this section  is derived the singlet commutation function
that will be used for all of the numerical results below.
In Appendices~\ref{Sec:H2} and \ref{Sec:pair},
respective extensions beyond linear order
and to the analogous pair commutation function are given.

Consider a system of $N$ particles
subject to singlet and pair potentials,
\begin{equation}
U({\bf q}) =
\sum_{j=1}^N u^{(1)}({\bf q}_j)
+  \sum_{j<k}^N u^{(2)}(q_{jk}).
\end{equation}
A central pair potential is assumed,
$q_{jk} = |{\bf q}_j-{\bf q}_k|$.

In view of the position eigenfunctions, Eq.~(\ref{Eq:qrpr}),
one can draw a distinction between the eigenvalue ${\bf q}$ and
the representation coordinate ${\bf r}$.
The Dirac $\delta$-function means that ultimately
these are close to each other.
Accordingly,
the local potential felt by particle $j$
in configuration ${\bf q}$ can be written as
\begin{eqnarray} \label{Eq:ujr}
u_j({\bf r}_j|{\bf q}) & = &
u^{(1)}({\bf r}_j)
+ \frac{1}{2} \sum_{k=1}^N \!^{(k \ne j )} u^{(2)}(|{\bf r}_j-{\bf r}_k|)
\nonumber \\ & \approx &
u^{(1)}({\bf r}_j)
+ \frac{1}{2} \sum_{k=1}^N \!^{(k \ne j )} u^{(2)}(|{\bf r}_j-{\bf q}_k|).
\end{eqnarray}
The factor of one half is because each pair interaction
will be counted twice in the total potential
\begin{equation}
U({\bf r}|{\bf q}) =
\sum_{j=1}^N u_j({\bf r}_j|{\bf q}).
\end{equation}

The approximation in the
second line of Eq.~(\ref{Eq:ujr})
makes this a single particle potential
since the representation coordinates of particles
other than $j$ do not appear.
The mathematical justification for this approximation
is a little obscure
(but see the following discussion, \S~\ref{Sec:Discus}),
and instead one can judge the ansatz by the results that it produces.

Conceptually what follows can be carried out in any dimension.
But to simplify the notation and to convey the general idea,
the following analysis will be carried out for a one-dimensional system,
$q_1 < q_2 < \ldots < q_N$.

Maintaining the distinction between ${\bf r}$ and ${\bf q}$ given above,
the energy eigenfunctions of the full system
are simply products of single particle energy eigenfunctions,
\begin{equation}
\Phi_{\bf n}({\bf r}|{\bf q}) =
\prod_{j=1}^N  \phi_{j,n_{j}}(r_{j}) ,
\end{equation}
where
\begin{equation}
\phi_{j,n_{j}}(r_{j})
\equiv
\phi_{n_{j}}(r_{j}|{\bf q}_{/j})
=
\phi_{n_{j}}(r_{j}|q_{j-1},q_{j+1}) .
\end{equation}
In general the pair potential decays with increasing separation,
and so the total potential is dominated by the singlet potentials
and the nearest neighbor pair potentials.
Hence for simplicity the focus will be on the nearest neighbor
contributions to the commutation functions,
as signified in the final equality.

The single particle Hamiltonian operator is
\begin{eqnarray}
\hat{\cal H}_j^{(1)}(r_j) & = &
\frac{-\hbar^2}{2m} \partial_{r_j}^2
+ u^{(1)}(r_j)
 \\ \nonumber &&
+ \frac{1}{2}   u^{(2)}(r_j-q_{j-1})
+ \frac{1}{2}   u^{(2)}(q_{j+1}-r_j),
\end{eqnarray}
and the eigenvalue equation is
\begin{equation}
 \hat{\cal H}_j^{(1)}(r_j)  \phi_{j,n_{j}}(r_{j})
=
 E_{j,n_{j}} \phi_{j,n_{j}}(r_{j}) .
\end{equation}
The subscript $j$ on the energy operator, eigenfunction, and eigenvalue
signifies the dependence on the nearest neighbor configuration,
eg.\ $ E_{j,n_{j}} \equiv E_{n_{j}}(q_{j-1},q_{j+1})$.
It is straightforward to solve this single particle
eigenvalue equation numerically (see \S~\ref{Sec:Eigen}).
For concision, no explicit distinction
will be made below between interior and terminal wave-functions;
for the latter, simply ignore one of the fixed particle coordinates
and potential.

\comment{  
There are in fact two distinct singlet systems to be solved:
the interior system, $2 \le j \le N-1$,
and the terminal system, $j=1$ or $j=N$.
In the latter case the pair potential should simply be neglected
if it involves $q_0$ or $q_{N+1}$.
One needs a three-dimensional matrix
to store the set of terminal eigenfunctions,
and a four-dimensional matrix to store the set of interior eigenfunctions.
For a symmetric singlet potential, one can exploit mirror plane symmetry.
To abbreviate the following equations
it will be left to the reader to distinguish between terminal
and interior functions.
} 

This definition of the one-particle energy eigenfunctions
means that
the commutation function of the total system
can be written as the sum of singlet commutation functions,
\begin{eqnarray}
W({\bf q},{\bf p})
& = &
\sum_{j=1}^N w^{(1)}(q_{j},p_{j}|q_{j-1},q_{j+1}) .
\end{eqnarray}
This would be exact if the singlet potential
were the only potential in the system;
it is an approximation for the present case
of singlet and pair potentials.


From Eq.~(\ref{Eq:def-Wp}),
the singlet commutation function is given by
\begin{eqnarray} \label{Eq:tw2}
\lefteqn{
e^{-\beta {\cal H}^{(1)}(q,p|q',q'') }
e^{ w^{(1)}(q,p|q',q'') }
}  \\
& = &
\frac{1}{\langle q| p \rangle }
\sum_n
e^{-\beta E_n(q',q'')}
\phi_n(q|q',q'') \, \check \phi_n(p|q',q'')  .\nonumber
\end{eqnarray}
Here $\phi_n(q|q',q'') = \langle q| n,q',q'' \rangle$,
is the $n$th energy eigenfunction
for the effective potential $u(q|q',q'')$,
and $\check \phi_n(p|q',q'') \equiv \langle n,q',q''| p \rangle
= \int \mathrm{d}q\;   
e^{-qp/i\hbar} \phi_n(q|q',q'')^* $
is essentially its Fourier transform.

The structure of Eq.~(\ref{Eq:tw2}) makes
it convenient to define the combined singlet commutation function as
$\tilde w^{(1)}(q,p|q',q'')
\equiv w^{(1)}(q,p|q',q'') - \beta {\cal H}^{(1)}(q,p|q',q'')$.
The classical singlet Hamiltonian that appears here
and in the preceding equation
is
\begin{eqnarray}
\lefteqn{
{\cal H}^{(1)}(q,p|q',q'')
}  \\
& = &
\frac{p^2}{2m} + u(q|q',q'')
\nonumber \\  & = & \nonumber
\frac{p^2}{2m}
+ u^{(1)}(q)
+ \frac{1}{2} \left[ u^{(2)}(q,q') + u^{(2)}(q,q'')  \right]  .
\end{eqnarray}
The weighting of one half for the pair potentials
ensures that the classical Hamiltonian of the full system
is exactly
\begin{equation}
{\cal H}({\bf q},{\bf p})
=
U^\mathrm{nnn}({\bf q})
+ \sum_{j=1}^N {\cal H}^{(1)}(q_j,p_j|q_{j-1},q_{j+1}) .
\end{equation}
The interaction potential beyond nearest neighbors,
\begin{equation}
U^\mathrm{nnn}({\bf q})
=
\sum_{j=1}^{N-2} \sum_{k=j+2}^N u^{(2)}(q_j,q_{k}) ,
\end{equation}
has been included here
even though their direct influence
on the commutation function has been neglected.

With these results and suppressing for the moment
the symmetrization function,
the phase space weight is
\begin{equation}
e^{-\beta {\cal H}({\bf q},{\bf p}) }
e^{W({\bf q},{\bf p}) }
=
\frac{ e^{-\beta U^\mathrm{nnn}({\bf q})} }{ Z(\beta) }
\prod_{j=1}^N
e^{ \tilde w^{(1)}(q_j,p_j|q_{j-1},q_{j+1}) }.
\end{equation}

\subsubsection{Discussion}  \label{Sec:Discus}

In general
the effective local field is determined by two principles:
\begin{itemize}
\item
for a given configuration ${\bf q}$,
the total effective energy
equals the actual energy,
$U({\bf q}|{\bf q}) = U({\bf q})$,

\item
identical particles share identically. 
\end{itemize}
The first principle means that the fractional weights
of the interaction potential in each local field to which it contributes
must add up to one.
The second principle means that the fractional weight
 must be the same for each identical particle
that shares an interaction potential.
These two principles can only be satisfied with weight one half
for the pair potential contribution to the local field
for identical particles.
Using instead the actual local field for each particle
would violate the first principle.
These principles can be applied as well for three-body etc.\ potentials.
For particles with different masses,
the share of the effective local field felt by a particle
is inversely proportional
to its mass,
(ie.\ the field acting on particle $j$
from the pair potential with particle $k$ has fraction $m_k/(m_j+m_k)$).

\comment{ 
The eigenvalue example in Appendix~\ref{Sec:LocalPair} suggests
that the effective local field with weight 1/2 for the pair potential
is exact in the low temperature limit.
However it also indicates that the actual local field,
(weight 1 for the pair potential)
is exact in the high temperature limit.
One should not read too much into this result
as ultimately it is the commutation function,
not the eigenvalues and eigenfunctions, that has to be obtained accurately.
In this regard the following analysis
is more relevant than the results in Appendix~\ref{Sec:LocalPair}.
} 

The present effective local field approach is a little like the
Born-Oppenheimer method of quantum chemistry, in that certain
coordinates are fixed in the calculation
of the wave function for the remainder.
One difference is that that method uses
the actual local field for the electrons,
whereas the present method uses an effective local field.
A second difference is that there the electron energy
eigenvalues are fed into the nuclear Schr\"odinger equation
as an effective potential, whereas here the singlet wave functions
and singlet states for any one particle are fully independent
of the states of the other particles.
But perhaps the major difference between the two is that
the Born-Oppenheimer method seeks the energy eigenfunction
of the total system,
whereas the present method at the singlet level
only aspires to the single particle energy eigenfunction.
Above this is used in a one-particle quantum average
to obtain the commutation function that will subsequently
be used for classical many-body averages;
the combined singlet energy eigenfunctions don't
approximate the total system energy eigenfunction.

This last point is quite important in understanding
what has been achieved here.
Because of the first principle,
$U({\bf q}|{\bf q}) = U({\bf q})$,
the total singlet Hamiltonian, $\hat {\cal H}({\bf r}|{\bf q})
= \sum_j  \hat{\cal H}_j^{(1)}(r_j)$,
has the same functional form as the actual total Hamiltonian operator
at ${\bf r} = {\bf q}$,
$\hat {\cal H}({\bf q}|{\bf q}) = \hat {\cal H}({\bf q})$.
But whereas in representation coordinates
$\hat {\cal H}({\bf r}|{\bf q}) \Phi_{\bf n}({\bf r}|{\bf q})
= E_{\bf n}({\bf q}) \Phi_{\bf n}({\bf r}|{\bf q})$,
with $E_{\bf n}({\bf q}) = \sum_j E_{j,n_j}$,
this does not hold for the position configuration itself,
$\hat {\cal H}({\bf q}) \Phi_{\bf n}({\bf q}|{\bf q})
\ne E_{\bf n}' \Phi_{\bf n}({\bf q}|{\bf q})$,
where $E_{\bf n}'$ could be a constant or $E_{\bf n}({\bf q})$.
In other words, this ansatz does \emph{not} give
the energy eigenfunctions of the total system.

Conversely, if $\Psi_{\bf n}({\bf r})$
is an eigenfunction of the total system Hamiltonian,
$\hat {\cal H}({\bf r}) \Psi_{\bf n}({\bf r})
= E_{\bf n}'' \Psi_{\bf n}({\bf r})$,
then one also has that
\begin{eqnarray}
\hat {\cal H}({\bf r}|{\bf q}) \Psi_{\bf n}({\bf r})
& = &
[ E_{\bf n}'' - U({\bf r}) + U({\bf r}|{\bf q}) ] \Psi_{\bf n}({\bf r})
\nonumber \\ & = &
E_{\bf n}'' \Psi_{\bf n}({\bf r}),
\;\; {\bf q} = {\bf r} .
\end{eqnarray}
Notice how the effective local field is essential to this result,
$U({\bf r}|{\bf r}) = U({\bf r})$;
the result would not hold for the actual local field
(because then $U({\bf r}|{\bf r})\ne U({\bf r})$).
At the specific point ${\bf q} = {\bf r} $,
$\Psi_{\bf n}({\bf r})$ has the appearance of
also being an eigenfunction of the total singlet Hamiltonian operator.
However this is a little deceptive because at the quadratic level,
$ \hat {\cal H}({\bf r}|{\bf q})^2 \Psi_{\bf n}({\bf r})
\ne E_{\bf n}''^2 \Psi_{\bf n}({\bf r})$ at $ {\bf q} = {\bf r}$.
In this sense 
$\hat {\cal H}({\bf r}|{\bf q}) $ is a linear approximation to
$\hat {\cal H}({\bf r}) $ at $ {\bf q} = {\bf r}$.
(In Appendix~\ref{Sec:H2},
the singlet Hamiltonian operator exact to quadratic order
in the relevant exponential expansion is derived;
the cubic order term is also given.)

This linear interpretation has an important bearing
on the above derivation of the commutation function.
A sleight of hand has enabled the singlet energy eigenfunctions
to be used in Eq.~(\ref{Eq:def-Wp}),
namely  $\hat {\cal H}({\bf r})$ has been replaced by
$\hat {\cal H}({\bf r}|{\bf q})$ in the fundamental definition
of the commutation function, Eq.~(\ref{Eq:def1-Wp}),
\begin{equation} \label{Eq:expHrq}
e^{-\beta \hat{\cal H}({\bf r})}
\approx
1-\beta \hat{\cal H}({\bf r})
 =
1-\beta \hat{\cal H}({\bf r}|{\bf q})
\approx
e^{-\beta \hat{\cal H}({\bf r}|{\bf q})}.
\end{equation}
This replacement is essential
for $|{\bf n}\rangle \equiv \Phi_{\bf n}({\bf r}|{\bf q})$
to proceed from Eq.~(\ref{Eq:def1-Wp}) to Eq.~(\ref{Eq:def-Wp}).
These are a complete orthogonal set of eigenfunctions
of $\hat{\cal H}({\bf r}|{\bf q})$, not of  $\hat{\cal H}({\bf r})$.

In evaluating the merits of
the effective local field approach developed in this paper,
the question isn't whether
$\Phi_{\bf n}({\bf q}|{\bf q})$ is in some sense an approximation
to the total system energy eigenfunction $\Psi_{\bf n}({\bf q})$,
but rather whether $\hat {\cal H}({\bf r}|{\bf q})$ is a good
approximation to $\hat {\cal H}({\bf r})$ for the purposes
of calculating the commutation function.

Because the above derivation involves a linearization of the exponential,
it is strictly exact at high temperatures,
$\beta \rightarrow 0$.
The subsequent re-exponentiation has the virtue
of making the procedure applicable at lower temperatures,
because it effectively creates an infinite series
in powers of inverse temperature,
albeit with approximate coefficients.
One should not underestimate the worth of this re-exponentiation
in going beyond a strictly linear approach and in extending
its applicability to lower temperatures.
For example, amongst other contributions it gives
the exponential of the kinetic and potential energies
to form the classical Maxwell-Boltzmann factor,
which is essential in the exact phase space formulation at all temperatures.
It is shown in the derivation of the cubic term in Appendix~\ref{Sec:H2}
that exponentiating the linear term contributes
substantially to the exact higher order low temperature terms.

The efficiency of the linear formulation and resummation
can be seen quantitatively as follows.
Define the difference between the exact
and the linear Hamiltonian operator,
\begin{equation}
\Delta \hat{\cal H}
\equiv
\hat{\cal H}({\bf r}) -\hat{\cal H}({\bf r}|{\bf q})
=
U({\bf r}) -U({\bf r}|{\bf q})
\equiv
 U_\Delta .
\end{equation}
Obviously $ U_\Delta = 0$ at ${\bf r} = {\bf q}$.
One can write the exact Maxwell-Boltzmann operator as
\begin{eqnarray}
e^{ -\beta \hat{\cal H}({\bf r}) }
& = &
e^{ -\beta [\hat{\cal H}({\bf r}|{\bf q}) +  U_\Delta ]}
\nonumber \\ & = &
\sum_{n=0}^\infty
\frac{(-\beta)^n}{n!} [\hat{\cal H}({\bf r}|{\bf q}) + U_\Delta]^n .
\end{eqnarray}
A binomial expansion of the $n$th term in the series
has first term in each case
$\hat{\cal H}({\bf r}|{\bf q})^n$,
which will resum to $e^{ -\beta \hat{\cal H}({\bf r}|{\bf q}) }$,
as if $\Delta \hat{\cal H}= U_\Delta $ had been neglected entirely.
The remaining terms in the binomial expansion are of the form
$ U({\bf r}|{\bf q})^j  U_\Delta^k \nabla^l  U_\Delta^m $,
and more complicated.
But because $ U_\Delta = 0$ at ${\bf r} = {\bf q}$,
one sees that this is zero unless there are no factors of $ U_\Delta$
remaining after the various gradients have been evaluated.
So the only non-zero correction terms consist solely of factors
of powers of
$(\nabla  U_\Delta)\cdot \nabla$, $(\nabla^2  U_\Delta)$,
$(\nabla \nabla^2  U_\Delta )\cdot \nabla$ etc.,
(and also $U$),
and no factor of the form $ U_\Delta^k$.
Hence significantly many terms in the excess sum vanish;
for example only 14\% of the possible corrections
for the cubic term are non-zero.
This, as well as the fact that higher gradients of the potential
vanish rapidly at large separations,
explains why the non-linear resummation
of the linear formulation,
$e^{ -\beta \hat{\cal H}({\bf r}) }
\approx e^{ -\beta \hat{\cal H}({\bf r}|{\bf q}) }$,
can be expected to be accurate even at low temperatures.

This idea is exploited in Appendix~\ref{Sec:H2},
where the exact part of the procedure is extended to quadratic
and cubic order.

The difference between the present approach and the earlier
harmonic local field (then called mean field) approach
\cite{Attard18b,Attard19b,Attard18c}
is that here the actual effective local potential is used,
and the energy eigenvalues, eigenfunctions and commutation function
are obtained numerically.
In the earlier approach,
\cite{Attard18b,Attard19b,Attard18c}
a harmonic approximation was fitted to the
instantaneous effective local field
and the consequent exact analytic simple harmonic oscillator
energy eigenvalues, eigenfunctions, and commutation function
were used.
Both approaches are linear resummed in the above sense.

That earlier  harmonic local field theory was tested
against  exact results  for a harmonic crystal
at both the singlet and pair level.\cite{Attard19b}
For that system there are no anharmonic contributions,
and so it tests the effective local field
and linear level Hamiltonian themselves.
Those results show that the singlet theory is accurate
at both high and low temperatures,
even for temperatures where the ground state is dominant.
On the basis of these harmonic local field numerical results,\cite{Attard19b}
and the numerical results to be presented below,
one can be optimistic that the effective local field approximation is viable.

\subsection{Symmetrization Function} \label{Sec:Symm}

The paper preceding this one,\cite{Attard19c}
summarizing earlier work,\cite{STD2,Attard18a}
showed that the symmetrization function
could be written as the exponential
of the sum of loop symmetrization functions.
That result invoked the thermodynamic limit, $N \rightarrow \infty$.
In the present paper the theory will be applied to relatively small systems,
$N=4$ and $N=5$, which also happen to be at relatively low densities.
Under these circumstances one can neglect products of
symmetrization loops.
Two further approximations will be as well invoked:
trimer and higher loops will be neglected;
and for the dimer loops that remain,
only nearest neighbors will be permuted.

There is no compelling computational reason
for making these approximations,
since the full symmetrization function is relatively easy to evaluate.
But this approximate treatment of the symmetrization function
is arguably comparable to the nearest neighbor source particle
treatment of the singlet commutation function above.
In any case one hardly expects the neglected higher order terms
to make a measurable contribution in the cases treated below.

In detail
the symmetrization function is\cite{STD2,Attard18a,Attard19c,Attard19b}
\begin{eqnarray}
\lefteqn{
\eta^\pm({\bf p},{\bf q})
} \nonumber \\
& \equiv &
\frac{1}{\langle {\bf p} | {\bf q} \rangle }
\sum_{\hat{\mathrm P}} (\pm 1)^p \,
\langle \hat{\mathrm P} {\bf p} | {\bf q} \rangle
\nonumber \\ & = &
\frac{1}{\langle {\bf p} | {\bf q} \rangle }
\left\{
\langle {\bf p} | {\bf q} \rangle
\pm \sum_{j,k} \!' \langle \hat{\mathrm P}_{jk}{\bf p} | {\bf q} \rangle
+ \sum_{j,k,\ell} \!'  \langle \hat{\mathrm P}_{jk}
\hat{\mathrm P}_{k\ell}{\bf p}
| {\bf q} \rangle
\right. \nonumber \\ && \mbox{ } \left.
+ \sum_{j,k,\ell,m} \!\!\!'\;
\langle \hat{\mathrm P}_{jk}\hat{\mathrm P}_{\ell m} {\bf p}
 | {\bf q} \rangle
\pm \ldots
\right\}
\nonumber \\ & \approx &
1 + \sum_{j<k} \eta^{\pm(2)}_{jk}
\nonumber \\ & \approx &
1 + \sum_{j=1}^{N-1} \eta^{\pm(2)}_{j,j+1} .
\end{eqnarray}
The dimer loop here is
\begin{equation}
\eta^{\pm(2)}_{jk}
=
\pm  e^{-{\bf p}_{ jk }  \cdot {\bf q}_{jk} /i\hbar} .
\end{equation}
For the present one dimensional problem,
${\bf p}_{ jk }  \cdot {\bf q}_{jk} = [p_j-p_k][q_j-q_k]$.

\subsection{Computer Algorithm} 

The computational problem is split into three programs:
first, obtain the one particle energy eigenvalues and eigenfunctions
for the fixed one-  and two-particle systems
(ie.\ terminal and interior variable particle);
second, construct the corresponding
singlet commutation functions;
and third,
perform a Monte Carlo simulation in classical phase space
with the commutation function and symmetrization function.

\subsubsection{Energy Eigenstates} \label{Sec:Eigen}

The one-particle systems have the potential,
\begin{equation}
U(r|q',q'') = u^{(1)}(r)
+ \frac{1}{2} \left[ u^{(2)}(r-q') + u^{(2)}(q''-r) \right].
\end{equation}
(For the terminal case one of the pair potentials is absent.)
The singlet potential is that of a simple harmonic oscillator,
\begin{equation}
u^{(1)}(r) = \frac{1}{2} m \omega^2 r^2,
\end{equation}
and the pair potential is Lennard-Jones,
\begin{equation}
u^{(2)}(r-q')=  \varepsilon \left[
\frac{r_\mathrm{e}^{12}}{(r-q')^{12}}
- 2
\frac{r_\mathrm{e}^{6}}{(r-q')^{6}}
\right] .
\end{equation}

The wave function is obtained on a three-dimensional grid
for $q$, $q'$, and $q''$,
with a fourth dimension for the energy level $n$.
One has
$q'+r_\mathrm{min} \le r \le q''-r_\mathrm{min}$,
with the fixed particles in the range
$-L/2 \le q' \le L/2-r_\mathrm{e} $,
and
$-L/2+r_\mathrm{e} \le q'' \le L/2 $.
Typically, $r_\mathrm{min} = 0.75 r_\mathrm{e} $
and $L = 10 r_\mathrm{e}$.

The $n$th energy eigenvalue $E_n(q',q'')$
and eigenfunction $\psi_n(r|q',q'')$
are obtained by minimizing the energy expectation value
$\langle n | \hat{\cal H}| n \rangle$
using a Gram-Schmidt procedure to ensure that
$\langle n | k \rangle = 0 $, $k=1,2,\ldots, n-1$.
An iterative procedure is used
where the change in wave function
is proportional to the force at each grid point.
A three-point finite difference formula is used for the
Laplacian for the kinetic energy,
with the gradient of the wave function at the terminal grid points
being used as part of the minimization procedures.
(A limited check was made using the derivative of the core asymptote
derived in Appendix~\ref{Sec:psi-core},
with no statistically significant effect.)
The first fifty singlet energy states are obtained in each case.
Roughly speaking,
this equates to on the order of $N^{50}$ energy levels for the full system.

The numerical procedures
are checked for the case of the simple harmonic oscillator
where the  energy eigenvalues
and eigenfunctions are known analytically.
The error in the 60th simple harmonic oscillator energy eigenvalue
is 1--3\%.
No numerical problems are experienced
with the Lennard-Jones potential,
but in this case there is no benchmark to test against.

The interior wave function is saved
as a matrix $\phi(j,n,j',j'')$
on a
$100 \times 50 \times 50  \times 50  $ grid.
To obtain the entire set of energy eigenfunctions
takes about 18 hours on an elderly personal computer.
No effort was made to optimize the process
because this is a once only computation
as the  wave functions are independent of temperature
and of the size of the ultimate system.

\subsubsection{Commutation Functions} 

The singlet commutation function
is constructed according to Eq.~(\ref{Eq:tw2}).
The Fourier integral is evaluated directly rather than by fast techniques.
Typically 100 momentum points are used on a uniform grid
up to a maximum kinetic energy $\beta {\cal K} = 50$.
The singlet commutation function
has to be calculated once for each temperature explored.
This computation is straightforward,
taking less time
than the several minutes spent writing the 412\,MB
output file to disc.

\subsubsection{Monte Carlo Simulations} 

The standard Metropolis algorithm is used,
with the umbrella weight being the real part
of the total combined commutation function,
including the beyond nearest neighbor potential contributions.
The imaginary contribution is added for taking averages,
and the commutation part is subtracted for taking classical averages.

It is found that at lower temperatures
it is more statistically efficient
to use umbrella sampling as above
than umbrella sampling using the classical Maxwell-Boltzmann weight,
or broader variant thereof.

The  commutation functions for a configuration
are obtained by linear interpolation
of the combined singlet commutation functions that are stored on the
3- or 4-dimensional grids.
This is more accurate and reliable
than interpolation of the commutation functions directly,
as a delicate cancelation has to occur
between some rather large numbers in the core region

Each simulation for a given temperature and system size
produces 6 averages of any given quantity:
either classical or quantum,
combined with either distinguishable particles, or bosons, or fermions.
A typical simulation for $N=4$ particles takes about 35 minutes on the
aforementioned personal computer,
for about a 1\% statistical error at the 96\% confidence level
of the average total energy for distinguishable particles.
The same set of energy eigenfunctions
are used for simulations for a dozen different temperatures
and for $N=4$ and $N=5$,
with several repeated with different simulation parameters.

%
\section{Results}
\setcounter{equation}{0} \setcounter{subsubsection}{0}
%

The numerical result reported here are
for a one dimensional system
with the simple harmonic oscillator singlet potential
and Lennard-Jones pair potential given above.
Quantitative comparison is made with the benchmark results
obtained by Hernando and Van\'i\v cek,\cite{Hernando13}
and following them the so-called de Broglie wave length
is fixed at $\Lambda_\mathrm{dB}
\equiv 2^{1/6}\hbar / r_\mathrm{e}\sqrt{m \varepsilon}
= 0.16$,
and the frequency is fixed by
$\omega r_\mathrm{e} \sqrt{m /\varepsilon} = 1/2$.
These mean that
$\varepsilon/\hbar \omega
= 14.03$.

\subsection{Commutation Effects on the Phase Space Weight}

\begin{figure}[t!]
\centerline{
\resizebox{8cm}{!}{ \includegraphics*{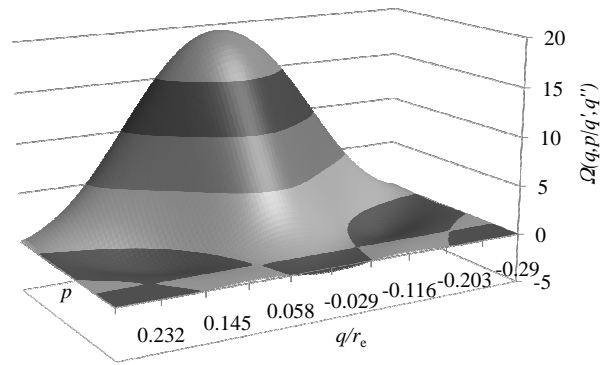} } }
\caption{\label{Fig:Omega-3d}
Phase space singlet weight,
$\mbox{Re } \Omega(q,p|q',q'')$,
for $q'' = -q' = r_\mathrm{e}$
and $\beta \hbar \omega = 0.5$.
The momentum depth axis ranges from 0 (back)
to $\beta p^2/2m = 50$ (front).
The darker squares on the checkered floor
have negative weight.
}
\end{figure}

Figure~\ref{Fig:Omega-3d} shows
the real part of the phase space weight
for a  single particle with fixed neighbors,
$\Omega^{(1)}_\mathrm{re}(q,p|q',q'') =
e^{-\beta{\cal H}^{(1)}(q,p|q',q'')}$
$e^{ \, w^{(1)}_\mathrm{re}(q,p|q',q'')}
\cos w^{(1)}_\mathrm{im}(q,p|q',q'')$.
It can be seen that this is a local maximum along the line $p=0$,
and, for symmetrical fixed particles,
$q'' = -q' $, along the line $q=0$.
The singlet phase space weight goes smoothly to zero
as $|p| \rightarrow \infty$
and as $q \rightarrow q'$ or $ q''$.

In the floor region of Fig.~\ref{Fig:Omega-3d},
$ |\mbox{Re } \Omega^{(1)}(q,p|q',q'')| \ll 1$,
small amplitude oscillations can be observed,
with the phase space weight being negative in places.
Such behavior is forbidden in classical statistical mechanics
and in classical probability theory,
but is allowed for quantum probabilities.
Indeed, quantum probabilities are complex.
It can be shown that if the phase space weight
is integrated with respect to position,
then the consequent momentum density is real and non-negative.
And if the phase space weight
is integrated with respect to momentum,
then the consequent position density is also real and non-negative.
The negative values
that appear periodically in the real part of the phase space weight
appear to be an actual consequence of quantum statistical mechanics.
It is \emph{possible} that they could be a numerical artefact
of the present grid and iteration procedures,
but this seems unlikely as they have been observed
for different parameter choices.
In either case the magnitude of the negative parts
is much smaller than the peak value of the density,
and so they appear to have negligible effect on the statistical averages.

\begin{figure}
\centerline{
\resizebox{8cm}{!}{ \includegraphics*{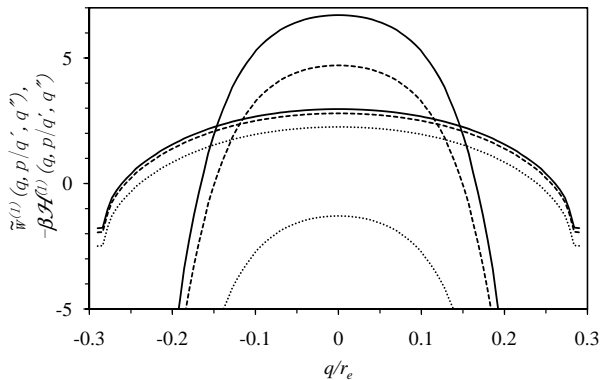} } }
\caption{\label{Fig:tw1-q}
Real part of the combined singlet commutation function
$ \mbox{Re } \tilde w^{(1)}(q,p|q',q'')$
(broad curves),
and the classical Maxwell-Boltzmann singlet exponent
$- \beta{\cal H}^{(1)}(q,p|q',q'')$ (narrow curves)
for $q'' = -q' = r_\mathrm{e}$
and $\beta \hbar \omega = 0.5$.
The momentum corresponds a kinetic energy of
$\beta p^2/2m =$ 0 (solid curves),
2 (dashed curves),
and 8 (dotted curves).
}
\end{figure}

Figure~\ref{Fig:tw1-q}
compares the exponent of the singlet phase space density,
the real part of the combined singlet commutation function,
$ \tilde w^{(1)}_\mathrm{re}(q,p|q',q'')
= w^{(1)}_\mathrm{re}(q,p|q',q'') - \beta{\cal H}^{(1)}(q,p|q',q'')$,
with the exponent of the classical singlet Maxwell-Boltzmann distribution,
$-\beta{\cal H}^{(1)}(q,p|q',q'')$.
As in the preceding figure the source particles are fixed at
$q'=-r_\mathrm{e}$ and  $q''=r_\mathrm{e}$;
the abscissa ranges over
$[r_\mathrm{min} + q', q''-r_\mathrm{min}]$,
with $r_\mathrm{min} = 0.75 r_\mathrm{e} $.
It can be seen that  accounting for the non-commutativity
of the position and momentum operators
reduces the range of the  exponent
(ie.\ broadens it and lowers its peak)
compared to the classical case.
This makes intuitive sense in that quantum particles
must be less localized than their classical counterparts.
That $ \tilde w^{(1)}_\mathrm{re} \gg -\beta{\cal H}^{(1)}$
toward the extremities, $ q \rightarrow q' $ or $q''$,
indicates that the quantum particles are more likely
to penetrate more deeply into the repulsive core region
than classical particles.
This suggests that one should not make the low separation cut-off,
$r_\mathrm{min}$, too large.
That the combined commutation function
is much smaller in magnitude than the Maxwell-Boltzmann exponent here
indicates the amount of cancelation that is required in the core region,
and hence the necessity of an accurate grid interpolation scheme here.
This core cancelation evident in Fig.~\ref{Fig:tw1-q}
is consistent with the behavior of the energy eigenfunctions
and the singlet commutation function derived in Appendix~\ref{Sec:psi-core}.

Figure~\ref{Fig:tw1-q} also shows
that the decrease in the exponent with increasing momentum
is much less marked in the quantum case than in the classical case.
This means that the quantum system can readily access much higher values
of momentum than are classically predicted.
This is presumably a manifestation of the Heisenberg uncertainty relation.

\subsection{Symmetrization Effects for Bosons and Fermions}

\begin{table}
\caption{ \label{Tab:BH0+-}
Average energy $\langle  {\cal H} \rangle /\hbar \omega$,
for $N=5$ distinguishable particles, bosons, and fermions.
Twice the standard error on the mean 
for the final digits is shown in parentheses.
}
\begin{tabular}{cccc}
\hline
$\beta \hbar \omega $ & 
Dist. & Bosons  & Fermions \\
\hline
0.8 & 
-0.4857(54) & -0.514(95) & -0.505(82) \\
0.9 & 
-0.6233(65) & -0.648(100) & -0.556(86) \\
1.0 & 
-0.7091(63) & -0.756(83) & -0.752(70)  \\
1.2 & 
-0.7544(62) & -0.735(61) & -0.761(49)  \\
1.5 & 
-0.8406(80) & -0.860(56) & -0.812(47)  \\
\hline
\end{tabular} \\
\end{table}

In these systems it is difficult to measure effects of symmetrization
that are statistically significant.
Nevertheless from the results in Table~\ref{Tab:BH0+-}
one might be persuaded that the average energy
for bosons is most likely less than that for fermions.
Conversely,
the results are not inconsistent with the hypothesis
that the symmetrization of the wave function
has no effect on the average energy.


Earlier exact and  harmonic local field
results for a harmonic crystal
showed that symmetrization effects were small,
and that the average energy for bosons could be greater than
or less than that of fermions, depending upon the temperature.
\cite{Attard19b}
In Ref.~[\onlinecite{Attard19a}]
exact calculations showing non-monotonic change in energy difference
between bosons and fermions were attributed to two competing effects:
with decreasing temperature the thermal wavelength increases,
whereas the overlap of adjacent singlet wave-functions decreases
as the particles are more closely confined to their lattice positions.

\subsection{Benchmark Tests}

\begin{figure}[t!]
\centerline{
\resizebox{8cm}{!}{ \includegraphics*{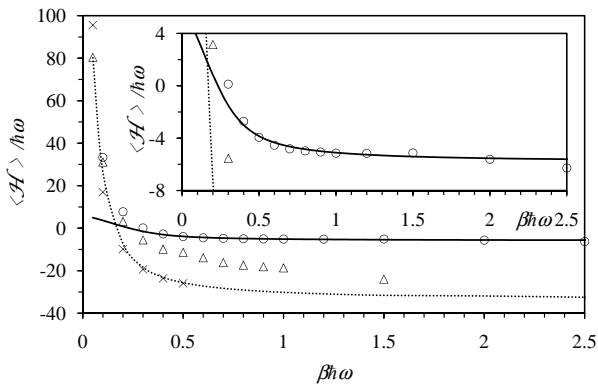} } }
\caption{\label{Fig:<H>4}
Average energy as a function of inverse temperature for N=4
distinguishable particles.
The circles result from Monte Carlo simulations
with the present singlet commutation function
with nearest neighbor local field.
The triangles are the earlier mean field simple harmonic oscillator
approach,\cite{Attard18c}
the crosses are a high temperature expansion,\cite{Attard19b}
the dotted curve is the classical result,
and the solid curve are benchmark results
derived from the 50 energy eigenvalues
given in Ref.~[\onlinecite{Hernando13}],
shifted vertically to coincide with the present results
at intermediate temperatures.
The statistical error is less than the symbol size.
}
\end{figure}

Figure~\ref{Fig:<H>4}
compares the present prediction for the average energy
for a system of $N=4$ distinguishable particles
with the  benchmark results derived from
Ref.~[\onlinecite{Hernando13}].
Unfortunately there is an unknown shift in the published energy eigenvalues.
\cite{Hernando13}
To make the comparison in the figure,
the average energy derived from the published energy eigenvalues
has been shifted vertically
to coincide with the present theory at intermediate temperatures,
where the present theory is believed to be the most reliable.
This of course limits the utility of the benchmarks
in absolute terms.
Nevertheless the slope and curvature of the present results
agree with those of the benchmark results at intermediate temperatures,
$0.3 \alt \beta \hbar \omega \alt 2 $
(inset),
which tends to confirm the quantitative validity of the present results.

One can estimate the ground state energy
in the local field approach gives as
\begin{eqnarray}
E_0^\mathrm{loc}
& \approx &
- (N-1) \varepsilon
+
\frac{(N-2)\hbar}{2} \sqrt{ \omega^2 + \omega_\mathrm{LJ}^2  }
\nonumber \\ & & \mbox{ }
+
\hbar \sqrt{ \omega^2 + \omega_\mathrm{LJ}^2 /2  } .
\end{eqnarray}
Here $\omega_\mathrm{LJ} = \sqrt{ u_\mathrm{LJ}''(r_\mathrm{e})/m}
=16.97 \omega$    
for the present parameters.
This gives $E_0^\mathrm{loc} /\hbar\omega = -13.1$ for $N=4$.
(The simulations using the second grid described below give
$\langle {\cal H} \rangle /\hbar\omega =$  $ -5.922 \pm 0.032 $
at $\beta \hbar\omega = 2$,
and $ -7.760 \pm 0.030$ at 5,
which linearly extrapolate to a ground state of $-8.99$.)
One can see in the inset to Fig.~\ref{Fig:<H>4}
that, after flattening at intermediate temperatures,
the average energy given by the present theory starts
to decrease toward this ground state as the temperature is further decreased.
This appears to underestimate the benchmark results
that have been shifted to the present intermediate temperature results,
with which shift they give a ground state of $E_0/\hbar\omega=-5.69$.
It is not completely clear but the present underestimate
is possibly an artefact of the singlet, linear approximation.
(The harmonic local field results,
which is in the same sense linear, underestimated
the exact ground state of a harmonic crystal at low temperatures,
and the error was about halved in going from the singlet to the
pair level.\cite{Attard19b})
In any case one can see
that including quantum effects
makes a very great difference
to the average energy at low temperatures;
the classical prediction is
$\langle {\cal H} \rangle_\mathrm{cl}/\hbar\omega = -33.3$ at absolute zero.

The local field results in Fig.~\ref{Fig:<H>4} were obtained
with an interior singlet commutation function
$w^{(1)}(q_j,p_j|q_{j-1},q_{j+1})$
on a $100 \times 100 \times 50 \times 50 $ grid,
with system length $L = 10 r_\mathrm{re}$.
As a test, some results were also obtained on a
$110 \times 100 \times 60 \times 60 $ grid with $L = 9 r_\mathrm{re}$.
For $\beta\hbar\omega = 2.0$,
the first grid gave $ \langle {\cal H} \rangle/\hbar\omega =$
$-5.615 \pm  0.029$ and the second gave $-5.922 \pm  0.032$.
At $\beta\hbar\omega = 0.7$,
the respective results were
$-4.813 \pm  0.017$ and $-4.963 \pm  0.019$,
and at $\beta\hbar\omega = 0.4$,
the respective results were
$ -2.707 \pm 0.014$ and $-2.976 \pm 0.015$.
The grid dependence is evidently small.


At high temperatures the present method is exact,
as can be seen by the agreement with the classical results
in Fig.~\ref{Fig:<H>4}.
The failure of the data derived from Ref.~[\onlinecite{Hernando13}]
in this regime is due to the fact
that they are based on only 50 energy eigenvalues for the full system.
The present calculations also obtained 50 energy eigenvalues,
but these were for a single particle system (with two fixed particles
providing the local field),
and they were used to construct the singlet commutation function.
Since the total commutation function is the sum of the latter,
(equivalently,
the total wave function is the product of singlet wave functions)
the present method effectively characterizes
the full system with $4^{50}$ energy levels.

It can be seen in Fig.~\ref{Fig:<H>4}
that the mean field, simple harmonic oscillator theory,
which would be better named a local field theory in harmonic approximation,
correctly gives the high temperature classical limit,
but disagrees with the present local field approach
at intermediate and low temperatures.
This, presumably, indicates that the anharmonic contributions
to the local potential field experienced by a Lennard-Jones particle
becomes increasingly significant as the temperature is decreased.
In this case it is better to obtain the exact energy eigenfunctions
for the actual local field
rather than to use the simple harmonic oscillator energy eigenfunctions
for the quadratic expansion of the local field.
(In higher dimensions,
the central limit theorem suggests
that the harmonic approximation would be more accurate.)

\begin{figure}[t!]
\centerline{
\resizebox{8cm}{!}{ \includegraphics*{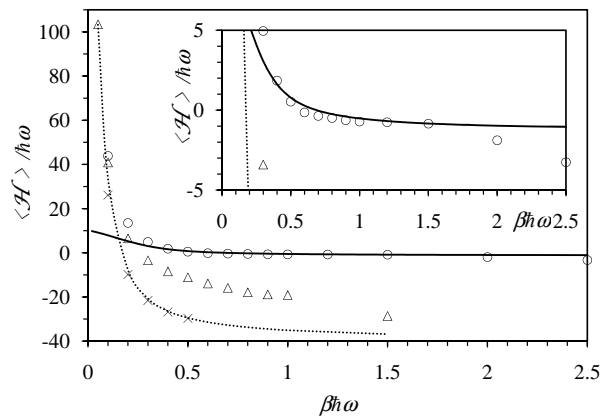} } }
\caption{\label{Fig:<H>5}
Same as the preceding figure but for $N=5$.
}
\end{figure}

Figure~\ref{Fig:<H>4} also includes the result
of a high temperature expansion for the commutation function.
\cite{Attard19b}
Although the terms in this expansion are formally exact,
it can be seen that the result is practically indistinguishable
from the classical result on the scale of the figure.
Given the complexity of the terms
and the brute force nature of the expansion,
this does not appear to be a promising approach
for the quantitative description of quantum effects
in terrestrial condensed matter.

Figure~\ref{Fig:<H>5} shows the average energy
for $N=5$ distinguishable particles.
Again it can be seen (inset)
that the slope and curvature of the energy curve
given by the present theory
agree with that derived from the eigenvalue data
of Ref.~[\onlinecite{Hernando13}]
in the intermediate temperature regime,
$0.4 \alt \beta \hbar \omega \alt 1.5 $.
The present two neighbor  local field approach,
the harmonic local field approach (ie.\ mean field),
and the high temperature expansion,
all go over to the classical limit at high temperatures.
The quadratic approximation to the local field
again appears inadequate at lower temperatures.
The ground state formula given above gives
$E_0^\mathrm{loc} /\hbar\omega = -18.6$ for $N=5$,
and the classical result  at absolute zero is
$\langle {\cal H} \rangle_\mathrm{cl}/\hbar\omega = -38.6$.
The present result again appears to underestimate
the shifted benchmark result,  $E_0/\hbar\omega=-1.13$.

Its worth mentioning the earlier  mean field (ie.\ harmonic local  field)
results for a quantum harmonic crystal
for which the exact analytic results are known.\cite{Attard19a}
At the relatively low temperature of
$\beta \hbar \omega_\mathrm{LJ} = 10$,
the singlet  harmonic mean field approach gave
$\langle {\cal H} \rangle /\hbar \omega_\mathrm{LJ}
= 3.116 \pm .002$
and the pair mean field gave $3.305 \pm .001$,
to be compared to the exact classical result of 0.4,
and the exact ground state energy of 3.385.\cite{Attard19b}
These show that the local field approach
is quite capable of giving accurate results even
at low temperatures where the quantum ground state is dominant.
This is perhaps surprising since the local field Hamiltonian operator
used in the harmonic calculations (and also here)
is strictly valid only up to the linear term
in a high temperature expansion.
Apparently the approximate non-linear terms induced by the re-exponentiation
produce accurate results also at low temperatures.

\begin{figure}[t!]
\centerline{
\resizebox{8cm}{!}{ \includegraphics*{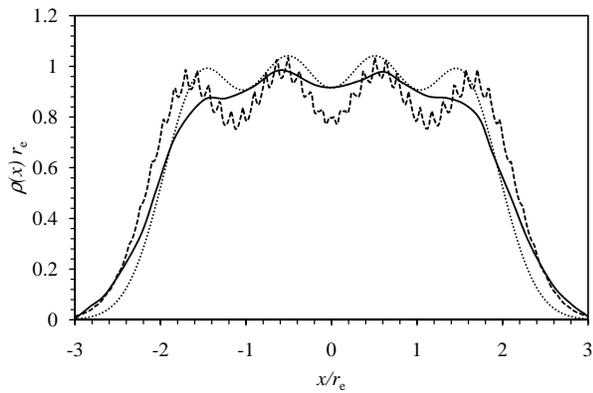} } }
\caption{\label{Fig:rho.4}
Density profile at $\beta \hbar \omega = 0.4$
for $N=4$ distinguishable particles.
The dashed curve is the present theory,
the dotted curve is the classical prediction,
and the full curve are the benchmark results
given  by Hernando and  Van\'i\v cek.\cite{Hernando13}
}
\end{figure}

Density profiles at  $\beta \hbar \omega = 0.4$
are shown in Fig.~\ref{Fig:rho.4}.
All three approaches give a density
symmetrically localized about the mid-plane,
as one would expect for the simple harmonic oscillator applied potential.
All approaches also predict molecular-sized oscillations
in the density profile,
which is indicative of a relatively close-packed system.
The outer peaks are less well-defined in the benchmark results
than in the classical or present singlet local field approaches.

The high frequency oscillatory fine structure
evident in the profile given by the present theory
is statistically significant.
It is not an artefact of the grids used to collect the density profile
or to store the singlet commutation function.
On the scale of the figure no difference would be seen
for distinguishable particles, bosons, or fermions.

\begin{figure}[t!]
\centerline{
\resizebox{8cm}{!}{ \includegraphics*{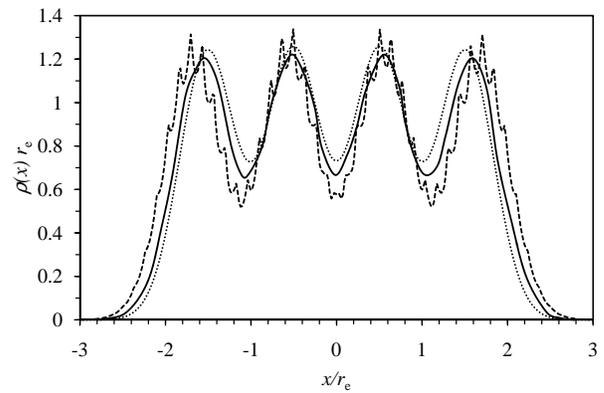} } }
\caption{\label{Fig:rho.7}
Same as the preceding figure but for $\beta \hbar \omega = 0.7$.
}
\end{figure}

Figure~\ref{Fig:rho.7}
shows the density profiles at a lower temperature,
$\beta \hbar \omega = 0.7$.
In this case the structure is much more well-defined,
and the departure from classical close-packing
is relatively small.
It is remarkable just how much of the density structure
is due to purely classical considerations
even at this low temperature.
Again the high frequency fine structure in the present density profile
is statistically significant
and it appears robust with respect to changes in the various grids
and particle statistics used in the simulations.

%
\section{Conclusion}
\setcounter{equation}{0} \setcounter{subsubsection}{0}
%

What has been obtained in this paper 
is an effective one-particle Hamiltonian operator
that can  be used ultimately to obtain properties
of the whole many-particle system.
What has \emph{not} been done is to obtain the energy eigenfunction
of the whole system (see \S~\ref{Sec:Discus}).

The singlet eigenfunctions $ \phi_{n_j}(r_j|{\bf q}_{/j})$
and eigenvalues $ E_{n_j}({\bf q}_{/j})$
are useful for obtaining certain one-particle quantum averages,
including the singlet commutation function,
$w^{(1)}(q_j,p_j|{\bf q}_{/j})$.
This function can be applied in classical phase space
to obtain the statistical average of many-body phase functions.
The phase space formulation
is an exact formulation of quantum statistical mechanics.
However basing the commutation function
on only one-particle eigenfunctions
is obviously an approximation to the full function.
But two points are clear:
first one can obtain at some level of accuracy
the full system quantum average without
obtaining the full system eigenfunction.
And second,
there are ways to systematically improve the reliability of the
quantum averages.

One method of improvement is given in Appendix~\ref{Sec:H2},
where temperature-dependent terms are added
to the singlet Hamiltonian operator,
which makes the singlet commutation function
exact to quadratic and cubic order in inverse temperature.
It would appear that a sufficiently large number of terms in this series
would give exact phase space results at any given temperature
even though they are based only on singlet eigenfunctions.

A second method of improvement is to proceed
to the pair, triplet etc.\ level
again using an effective local field
that is parameterized by the neighboring particles' positions.
A pair formulation is given in Appendix~\ref{Sec:pair}.
Again a sufficiently large number of terms in this many-body sequence
would yield exact phase space results at any given temperature
even without non-linear corrections
to the effective many-body Hamiltonian operator.
(Obviously the $N$-body linear effective Hamiltonian
is the same as the actual Hamiltonian operator of the $N$ particle system,
and the consequent linear effective eigenfunctions
would be actual eigenfunctions of the system.)

The singlet and pair formulation
of the commutation function derived in this paper
is not the same as the many-body effective potential expansion
derived in Ref.~[\onlinecite{Attard19c}].
The singlet commutation function used for the present numerical results
is really a three-body function;
more generally for $n$-fixed neighbors it is
an $(n+1)$-body effective potential.
The many-body expansion proposed in Ref.~[\onlinecite{Attard19c}]
has yet to be implemented or tested.

\comment{ 

The present paper has presented a practical algorithm
for computing the commutation function of a many particle system.
This function is necessary for quantum systems to be analyzed
in classical phase space.

Comparison of the present results with exact results
for the one-dimensional Lennard-Jones system
demonstrate once again the utility of the classical phase space approach
to quantum statistical mechanics.
The numerical results agree
with the classical results at high temperatures,
and with conventional quantum mechanical benchmarks at intermediate,
and they smoothly interpolate between the two.
The present results allow this conclusion
to be explicitly drawn for a fluid system of interacting particles,
which complements previous validation for
ideal, non-interacting systems,\cite{STD2}
and for a solid crystal of interacting particles.\cite{Attard19b}
The results demonstrate
that the present algorithm is feasible computationally,
both in terms of obtaining the commutation function
and in using it in a molecular simulation program.

Beyond the present  one-dimensional application,
there are  good prospects of applying the approach
to higher dimensional systems,
and of going beyond the singlet commutation function.
In a conceptual sense dimensionality plays no role
as far as the local field is concerned.
Undoubtedly there will be practical challenges in higher dimensions
in choosing the fixed neighbors
and in parameterizing and storing the local field,
but such difficulties should not prove insurmountable.

} 

The pair local field of Appendix~\ref{Sec:pair}
has, in harmonic approximation,
been tested against exact result for a harmonic crystal,
where it improved the accuracy of the singlet results.
\cite{Attard19b}
To obtain numerically the energy eigenvalues and eigenfunctions
more generally for a two-particle system with fixed local field
lies within current computational capabilities.
It is possible that the pair commutation function
captures correlations that are absent in the singlet formulation,
which could be relevant at, say, low temperatures, or high densities.

Finally, at a conceptual level, the present ansatz reconciles
the action at a distance inherent in quantum mechanics
with the local physics inherent in classical statistical mechanics.
Although there are specific quantum systems
that are coherent over macroscopic length scales,
the vast majority of terrestrial condensed matter systems are localized.
For these it seems strange that one should have to calculate
the wave function of the entire system
even though specific quantum effects are confined to much smaller regions.
The present approach shows that a viable alternative
is to obtain the few-particle eigenfunctions
in an effective local field due to fixed neighbors,
and to use these in a modified average in classical phase space
to describe the full quantum system.




\appendix



%
\section{Higher Order Corrections}
\label{Sec:H2}
\setcounter{equation}{0} \setcounter{subsubsection}{0}
\renewcommand{\theequation}{\Alph{section}.\arabic{equation}}
%

In \S~\ref{Sec:Discus}
it was shown that the effective local field
gave a singlet Hamiltonian operator
that coincided with the actual Hamiltonian operator
exactly to linear order.
This appendix derives the correction
to this singlet operator that makes it exact to quadratic order
in the expression for the commutation function.
Here and throughout
the gradient operator means a derivative with respect to ${\bf r}$,
and these are all performed \emph{before}
setting ${\bf r}$ equal to ${\bf q}$.

The exact and effective potentials
differ in the gradient of the pair potential.
For the pair part of the exact Hamiltonian one has
\begin{eqnarray}
\nabla^2 U({\bf r})
& = &
\sum_j \nabla_j^2 \sum_{k<l} u^{(2)}({\bf r}_k,{\bf r}_l)
\nonumber \\ & = &
\sum_{j,k}\!^{(j \ne k)}  \nabla_j^2  u^{(2)}({\bf r}_j,{\bf r}_k) .
\end{eqnarray}
Similarly, $ \nabla_j U({\bf r})
= \sum_{k}^{(k \ne j)}  \nabla_j  u({\bf r}_j,{\bf r}_k)$.
For the pair part of the  effective local potential one has
\begin{eqnarray}
\nabla^2 U({\bf r}|{\bf q})
& = &
\sum_j \nabla_j^2
\sum_{k=1}^N \frac{1}{2} \sum_l\!^{(l \ne k)} u^{(2)}({\bf r}_k,{\bf q}_l)
\nonumber \\ & = &
\frac{1}{2}\sum_j \sum_l\!^{(l \ne j)}
\nabla_j^2 u^{(2)}({\bf r}_j,{\bf q}_l) .
\end{eqnarray}
At ${\bf r} = {\bf q}$ this is half the previous expression.
Similarly, $ \nabla_j U({\bf r},{\bf q})
= \sum_{l}^{(l \ne j)}  \nabla_j  u^{(2)}({\bf r}_j,{\bf q}_l) /2$.

Now correct the linear singlet Hamiltonian operator
so that the linear expansion of the exponential
contains the currently missing quadratic part. That is
\begin{eqnarray}
-\beta^2 \hat \Delta^{(2)} 
& \equiv &
\frac{\beta^2}{2} \left\{
\hat{\cal H}({\bf r})^2 - \hat{\cal H}({\bf r}|{\bf q})^2
\right\}
 \\ & = &\nonumber
\frac{-\beta^2\hbar^2}{4m}
\frac{1}{2}\sum_{j,k}\!^{(k \ne j)}\nabla_j^2 u^{(2)}({\bf r}_j,{\bf q}_k)
 \\ & & \mbox{ }\nonumber
-
\frac{\beta^2\hbar^2}{2m}
\frac{1}{2}\sum_{j,k}\!^{(k \ne j)}\nabla_j u^{(2)}({\bf r}_j,{\bf q}_k)
\cdot \nabla_j .
\end{eqnarray}
Hence the temperature-dependent singlet Hamiltonian operator
that gives exact results at quadratic order
is
\begin{eqnarray}
\hat{\cal H}_j^{(1,2)}({\bf r}_j|{\bf q})
& = &
\hat{\cal H}_j^{(1)}({\bf r}_j|{\bf q})
+ \frac{\beta\hbar^2}{8m}
\sum_{k}\!^{(k \ne j)} \nabla_j^2 u^{(2)}({\bf r}_j,{\bf q}_k)
\nonumber \\ & & \mbox{ }
+ \frac{\beta\hbar^2}{4m}
\sum_{k}\!^{(k \ne j)} \nabla_j u^{(2)}({\bf r}_j,{\bf q}_k) \cdot \nabla_j
\nonumber \\ & = &
\frac{-\hbar^2}{2m} \nabla_j^2 + u_j({\bf r}_j|{\bf q})
+ \frac{\beta\hbar^2}{4m} (\nabla_j^2u^{(2)}_j({\bf r}_j|{\bf q}) )
\nonumber \\ & & \mbox{ }
+  \frac{\beta\hbar^2}{2m}
(\nabla_j u^{(2)}_j({\bf r}_j|{\bf q})) \cdot \nabla_j.
\end{eqnarray}
where $ u^{(2)}_j({\bf r}_j|{\bf q}) =
\sum_{k}\!^{(k \ne j)} u^{(2)}({\bf r}_j,{\bf q}_k)/2$
is the pair part only of the effective local field.
The eigenvalues $E_{j,n_j}^{(1,2)}({\bf q})$
and eigenfunctions  $\phi_{j,n_j}^{(1,2)}({\bf r}_j|{\bf q})$
of this Hamiltonian operator
can be used to construct a singlet commutation function
$w^{(1,2)}_j({\bf q}_j,{\bf p}_j)$,
the series of which gives the total commutation function of the system.
This extends the exact part of the approach in the text
from linear to quadratic order in inverse temperature.

\subsubsection{Cubic Correction}

With the exponent being written
$-\beta \hat{\cal H}^{(1,3)}({\bf r}|{\bf q}) \equiv
-\beta [ \hat {\cal H}^{(1,1)}({\bf r}|{\bf q})
+ \beta \hat \Delta^{(2)} + \beta^2 \hat \Delta^{(3)} ]$,
the cubic correction is given by
\begin{eqnarray}
\lefteqn{
- \beta^3 \hat \Delta^{(3)}
} \nonumber \\
& = &
\frac{-\beta^3}{3!}
\left[ \hat{\cal H}({\bf r})^3
-\hat{\cal H}^{(1,1)}({\bf r}|{\bf q})^3\right]
\nonumber \\ &  &  \mbox{ }
- \frac{\beta^2}{2!}
\left[ \hat{\cal H}^{(1,1)}({\bf r}|{\bf q})
\beta \hat\Delta^{(2)}
+
\beta \hat\Delta^{(2)} \;
\hat{\cal H}^{(1,1)}({\bf r}|{\bf q})
\right]
\nonumber \\ & = &
\frac{-\beta^3}{12}
\left\{ \rule{0cm}{0.4cm}
2(\nabla^2 \nabla^2 U_\Delta)
+ 8 (\nabla\nabla^2 U_\Delta) \cdot \nabla
+ 6(\nabla^2 U_\Delta)\nabla^2
\right. \nonumber \\ &  & \left. \mbox{ }
+ 8 (\nabla \nabla U_\Delta) : \nabla\nabla
+ 12 (\nabla U_\Delta) \cdot \nabla \nabla^2
\right. \nonumber \\ &  & \left. \mbox{ }
+ 8 (\nabla U_\Delta) \cdot (\nabla U)
+ 6 U (\nabla^2 U_\Delta)
+ 12 U ( \nabla U_\Delta) \cdot \nabla
\right. \nonumber \\ &  & \left. \mbox{ }
+ 4 (\nabla U_\Delta) \cdot (\nabla U_\Delta)
\rule{0cm}{0.4cm} \right\}
\nonumber \\ &  &
+ \frac{\beta^3}{12}
\left\{ \rule{0cm}{0.4cm}
3(\nabla^2  \nabla^2 U_\Delta)
+ 12 (\nabla  \nabla^2 U_\Delta) \cdot \nabla
+ 6( \nabla^2 U_\Delta) \nabla^2
\right. \nonumber \\ &  & \left. \mbox{ }
+ 6 U ( \nabla^2 U_\Delta)
+ 12 (\nabla\nabla U_\Delta) : \nabla \nabla
+ 12 (\nabla U_\Delta) \cdot \nabla \nabla^2
\right. \nonumber \\ &&  \left. \mbox{ }
+ 6 (\nabla U_\Delta) \cdot (\nabla U)
+ 6 U(\nabla U_\Delta) \cdot \nabla
\rule{0cm}{0.4cm} \right\}
\nonumber \\ & = &
\frac{\beta^3}{12}
\left\{ \rule{0cm}{0.4cm}
(\nabla^2 \nabla^2 U_\Delta)
+ 4 (\nabla\nabla^2 U_\Delta) \cdot \nabla
+ 4 (\nabla \nabla U_\Delta) : \nabla\nabla
\right. \nonumber \\ &  & \left. \mbox{ }
- 2 (\nabla U_\Delta) \cdot (\nabla U)
-6 U ( \nabla U_\Delta) \cdot \nabla
\right. \nonumber \\ &  & \left. \mbox{ }
- 4 (\nabla U_\Delta) \cdot (\nabla U_\Delta)
\rule{0cm}{0.4cm} \right\} .
\end{eqnarray}
Here $\nabla $ should be replaced by
$\sqrt{-\hbar^2/2m}\; \partial_{\bf r}$,
$U_\Delta \equiv U({\bf r}) - U({\bf r}|{\bf q})$,
and $U \equiv U({\bf r}|{\bf q})$.
One sees that there is substantial cancelation
between the exact and approximate third order terms
based on $\hat {\cal H}^{(1,1)}$ and $\hat\Delta^{(2)}$,
which means that the third order correction $\hat\Delta^{(3)}$
is smaller than if the linear and quadratic terms had not been exponentiated.
This is very strong evidence that re-exponentiation is valuable,
and that even at the linear level
the procedure already accounts for much of the low temperature contribution.

%
\section{Pair Theory}
\label{Sec:pair}
\setcounter{equation}{0} \setcounter{subsubsection}{0}
\renewcommand{\theequation}{\Alph{section}.\arabic{equation}}
%

Consider a system with singlet and pair potentials
\begin{equation}
U({\bf r}) =
\sum_{j=1}^N u^{(1)}({\bf r}_j)
+ \frac{1}{2} \sum_{j,k}\!^{(j \ne k)}  u^{(2)}({\bf r}_j-{\bf r}_k) .
\end{equation}
Assume three-dimensional space.

For any configuration ${\bf q}$,
one can relabel the particles
and group them into disjoint pairs, $\{2j,2j-1\}$.
The particles assigned to a pair
may vary from configuration to configuration.
Define the six-dimensional pair configuration as
${\bf Q}_j \equiv \{ {\bf q}_{2j},{\bf q}_{2j-1} \}$,
and similarly for the representation coordinate,
${\bf R}_j \equiv \{ {\bf r}_{2j},{\bf r}_{2j-1} \}$.

Define the pair-pair interaction potential as
\begin{eqnarray}
u^{(4)}({\bf R}_{j},{\bf R}_{k})
& \equiv &
u^{(2)}({\bf r}_{2j},{\bf r}_{2k})
+
u^{(2)}({\bf r}_{2j-1},{\bf r}_{2k})
 \\ && \mbox{ }\nonumber
+
u^{(2)}({\bf r}_{2j},{\bf r}_{2k-1})
+
u^{(2)}({\bf r}_{2j-1},{\bf r}_{2k-1}).
\end{eqnarray}
The singlet energies and the internal pair energies
can be included in the total energy by defining
\begin{equation}
u^{(2)}({\bf R}_{j})
\equiv
u^{(1)}({\bf r}_{2j}) + u^{(1)}({\bf r}_{2j-1})
+ u^{(2)}({\bf r}_{2j},{\bf r}_{2j-1}) .
\end{equation}

The total energy may now be written as
\begin{equation}
U({\bf R})
=
\sum_{j=1}^{N/2}  u^{(2)}({\bf R}_{j})
+
\sum_{j<k}^{N/2} u^{(4)}({\bf R}_{j},{\bf R}_{k}) .
\end{equation}
This has exactly the same functional form as the singlet theory,
with $u^{(2)}$ playing the role of the singlet potential
and $u^{(4)}$ that of the pair potential,
now in six- rather than three-dimensional space.

By analogy with the singlet formulation,
define the effective pair local potential as
\begin{eqnarray}
u^{(2)}_{j}({\bf R}_{j})
& \equiv &
u^{(2)}({\bf R}_{j}|{\bf Q}_{/j})
 \\ & = & \nonumber
u^{(2)}({\bf R}_{j})
+
\frac{1}{2} \sum_{k=1}^{N/2} \!^{(k \ne j)}
 u^{(4)}({\bf R}_{j},{\bf Q}_{k}).
\end{eqnarray}
With this the total effective local pair energy is
\begin{eqnarray}
\lefteqn{
U^{(2)}({\bf R}|{\bf Q})
\equiv
\sum_{j=1}^{N/2} u^{(2)}_{j}({\bf R}_{j})
}  \\ \nonumber
& = &
\sum_{j=1}^{N/2}
u^{(2)}({\bf R}_{j})
+
\frac{1}{2} \sum_{j=1}^{N/2} \sum_{k=1}^{N/2} \!^{(k \ne j)}
u^{(4)}({\bf R}_{j},{\bf Q}_{k}) .
\end{eqnarray}
Clearly,
$U^{(2)}({\bf Q}|{\bf Q}) = U({\bf Q})$.

The effective pair Hamiltonian operator for pair $j$ is
\begin{equation}
\hat {\cal H}_j^{(2)}({\bf R}_j)
=
\frac{-\hbar^2}{2m} \nabla_j^2 + u^{(2)}({\bf R}_{j}|{\bf Q}_{/j}) .
\end{equation}
The gradient operator is of course six-dimensional.
With this  Schr\"odinger's equation can be solved
to obtain the eigenvalues $E_{n_j}^{(2)}({\bf Q}_{/j})$
and eigenfunctions  $\phi_{n_j}^{(2)}({\bf R}_j|{\bf Q}_{/j})$.

The total effective Hamiltonian operator is
\begin{equation}
\hat {\cal H}^{(2)}({\bf R}|{\bf Q})
=
\sum_{j=1}^{N/2}
\hat {\cal H}_j^{(2)}({\bf R}_j) ,
\end{equation}
with eigenfunction
$ \Phi^{(2)}({\bf R}|{\bf Q})
= \prod_{j=1}^{N/2} \phi_{n_j}^{(2)}({\bf R}_j|{\bf Q}_{/j})$
and eigenvalue
$E_{\bf n}^{(2)}({\bf Q})
= \sum_{j=1}^{N/2} E_{n_j}^{(2)}({\bf Q}_{/j})$.

The effective pair
eigenvalues, $E_{n_j}^{(2)}({\bf Q}_{/j})$,
and eigenfunctions, $\phi_{n_j}^{(2)}({\bf Q}_j|{\bf Q}_{/j})$,
allow the pair commutation function to be obtained,
$w^{(2)}({\bf Q}_j,{\bf P}_j|{\bf Q}_{/j})$.
These give the total commutation function
as a sum over two-body  commutation functions.


One suspects that the pair theory
will be more accurate than the singlet theory.
This indeed turned out to be the case
for the linear numerical results already
obtained for the harmonic crystal.\cite{Attard19b}

%
\section{Core Asymptote} \label{Sec:psi-core}
\setcounter{equation}{0} \setcounter{subsubsection}{0}
\renewcommand{\theequation}{\Alph{section}.\arabic{equation}}
%

This appendix derives the behavior of the energy eigenfunction
in the  Lennard-Jones core region.
The full potential is
$u(r) = \varepsilon [ (r_\mathrm{e}/r)^{12} - 2 (r_\mathrm{e}/r)^{6} ]$,
 $r > 0$.
Take the energy eigenfunction function to be of the form
\begin{equation}
\psi(r) =
g(r) \exp \left\{ a (r_\mathrm{e}/r)^{5}  \right\} ,
\end{equation}
with $a =  -\sqrt{ {2m\varepsilon r_\mathrm{e}^2}/{25 \hbar^2} }$.
The energy eigenvalue equation is then
\begin{eqnarray}
\lefteqn{
E \psi(r)
}  \\
& = &
 \varepsilon [ (r_\mathrm{e}/r)^{12} - 2 (r_\mathrm{e}/r)^{6} ] \psi(r)
 - \frac{\hbar^2}{2m}
\left[  25 a^2 r_\mathrm{e}^{10} r^{-12} g(r)
\right. \nonumber \\ && \left. \mbox{ }
+ 30 a r_\mathrm{e}^{5} r^{-7} g(r)
- 10 a r_\mathrm{e}^{5} r^{-6} g'(r)
+ g''(r) \right] e^{a (r_\mathrm{e}/r)^{5} } .\nonumber
\end{eqnarray}
This gives
\begin{eqnarray}
\frac{-2mE}{\hbar^2}  g(r)
& = &
\frac{4m\varepsilon}{\hbar^2} (r_\mathrm{e}/r)^{6} g(r)
+ 30 a r_\mathrm{e}^{5} r^{-7} g(r)
\nonumber \\ &&  \mbox{ }
- 10 a r_\mathrm{e}^{5} r^{-6} g'(r) + g''(r) .
\end{eqnarray}
Insert
$g(r) = \sum_{n=3}^\infty g_n r_\mathrm{e}^{-n} r^n $,
equate coefficients of $(r/r_\mathrm{e})^n$,
and rearrange to obtain
\begin{eqnarray}
 g_{n+7}
 & = &
 \frac{1}{ 10 a (n+4) }
\left\{
\frac{4m\varepsilon r_\mathrm{e}^{2}}{\hbar^2} g_{n+6}
+ (n+2)(n+1)  g_{n+2}
\right. \nonumber \\ && \mbox{ }\left.
+ \frac{2mE r_\mathrm{e}^{2} }{\hbar^2} g_n
\right\} ,
\;\; n \ge -3,
\end{eqnarray}
with $g_n = 0 $, $n<3$.
This is homogeneous in $g_3$, 
so all coefficients can be rescaled
to  match the asymptote to the actual eigenfunction at a boundary.

In view of this asymptotic behavior
and  Eq.~(\ref{Eq:tw2}),
in the core region the combined singlet commutation function
must behave as
\begin{eqnarray}
\tilde w^{(1)}(q,p|q',q'')
& \equiv &
w^{(1)}(q,p|q',q'')
-\beta \frac{p^2}{2m}
-\beta u^{(1)}(q)
\nonumber \\ && \mbox{ }
- \frac{\beta}{2} \left[ u^{(2)}(q-q') + u^{(2)}(q-q'') \right]
\nonumber \\ & \sim &
-\sqrt{ \frac{2m r_\mathrm{e}^2 \varepsilon/2}{25 \hbar^2} }
\left(\frac{r_\mathrm{e}}{q-q'}\right)^5 ,
\end{eqnarray}
for $ 0 < q - q' \ll r_\mathrm{e}$.
Note the $\varepsilon/2$
due to the half weight for the pair potential in the local field.
The next term beyond the one shown is logarithmic.
To leading order this is independent of momentum and temperature,
and it gives a phase space weight
that vanishes in the core much more slowly
than that due to the Lennard-Jones potential itself.
This is qualitatively in agreement
with the results in Fig.~\ref{Fig:tw1-q}.


\begin{thebibliography}{99}


\bibitem{STD2}
P. Attard, \emph{Entropy Beyond the Second Law. Thermodynamics and
Statistical Mechanics for Equilibrium, Non-Equilibrium, Classical,
and Quantum Systems}, (IOP Publishing, Bristol, 2018).

\bibitem{Attard18a}
P. Attard, ``Quantum Statistical Mechanics in Classical Phase Space.
Expressions for
  the Multi-Particle Density, the Average Energy, and the Virial Pressure'',
arXiv:1811.00730 [quant-ph] (2018).

\bibitem{Wigner32}
E. Wigner,
``On the Quantum Correction for Thermodynamic Equilibrium'',
Phys.\ Rev.\ {\bf 40}, 749 (1932).

\bibitem{Kirkwood33}
J. G. Kirkwood,
``Quantum Statistics of Almost Classical Particles'',
Phys.\ Rev.\ {\bf 44}, 31 (1933).


\bibitem{Attard16}
P. Attard, ``Quantum Statistical Mechanics as an Exact Classical
Expansion with Results for Lennard-Jones Helium'',
arXiv:1609.08178v3 [quant-ph]  (2016). 




\bibitem{Attard18b}
P. Attard, ``Quantum Statistical Mechanics in Classical Phase Space.
Test Results for Quantum Harmonic Oscillators'', arXiv:1811.02032
(2018).  


\bibitem{Attard19b}
P. Attard,
``Quantum Monte Carlo in Classical Phase Space. Mean Field and Exact
  Results for a One Dimensional Harmonic Crystal'',
  arXiv:1904.10650  (2019). 

\bibitem{Attard18c}
P. Attard, ``Quantum Statistical Mechanics in Classical Phase Space.
III. Mean Field Approximation Benchmarked for Interacting
Lennard-Jones Particles'', arXiv:1812.03635 [quant-ph] (2018).


\bibitem{Attard19c}
P. Attard,
``Quantum Ornstein-Zernike Equation'',
arXiv:1908.06373v2 [quant-ph] (2019).

\bibitem{Hernando13}
A. Hernando and J. Van\'i\v cek,
``Imaginary-time nonuniform mesh method for solving the multidimensional
Schr\"odinger equation:
Fermionization and melting of quantum Lennard-Jones crystals'',
Phys.\ Rev.\ A {\bf 88}, 062107 (2013). arXiv:1304.8015v2 [quant-ph]
(2013).


\bibitem{Messiah61}
A. Messiah, \emph{Quantum Mechanics},
(North-Holland, Amsterdam, Vols I and II, 1961).

\bibitem{Attard19a}
P. Attard, ``Fermionic Phonons: Exact Analytic Results and Quantum
Statistical
  Mechanics for a One Dimensional Harmonic Crystal'',
arXiv:1903.06866 [quant-ph] (2019).




\comment{ 


\bibitem{Pathria72}
R. K. Pathria,
\emph{Statistical Mechanics},
(Pergamon Press, Oxford, 1972).

\bibitem{Hansen86}
J.-P. Hansen and I. R. McDonald
\emph{Theory of Simple Liquids},
(Academic Press, London, 1986).


\bibitem{TDSM}
P. Attard, \emph{Thermodynamics and Statistical Mechanics:
Equilibrium by Entropy Maximisation} (Academic Press, London, 2002).


\bibitem{Morita61}
T. Morita and K Hiroike,
``A New Approach to the Theory of Classical Fluids.
III ---General Treatment of Classical Systems---'',
Progr.\ Theor.\ Phys.\ {\bf 25}, 537--578  (1961).


\bibitem{Stell64}
G. Stell, ``Cluster Expansions for Classical Systems in Equilibrium'',
in \emph{The Equilibrium Theory of Classical Fluids},
(H. L. Frisch and J. L. Lebowitz, eds,
p.\ II:171--266, W. A. Benjamin, New York, 1964).


\bibitem{Attard92}
P. Attard,
``Pair Hypernetted Chain Closure for Fluids with Three-body Potentials.
  Results for Argon with the Axilrod-Teller Triple Dipole Potential.''
Phys.\ Rev.\ A {\bf 45}, 3659--3669 (1992).



\bibitem{Merzbacher70}
E. Merzbacher, \emph{Quantum Mechanics},
(Wiley, New York, 2nd ed., 1970).


\bibitem{QSM}
P. Attard, \emph{Quantum Statistical Mechanics: Equilibrium and
Non-Equilibrium Theory from First Principles}, (IOP Publishing,
Bristol, 2015).

\bibitem{Gasiorowicz74}
S. Gasiorowicz, \emph{Quantum Physics} (Wiley, New York, 1974).

\bibitem{Strauss68}
H. L. Strauss, \emph{Qunatum Mechanics: An Introduction} (Prentice
Hall, Englewood Cliffs, New Jersey, 1968).




\bibitem{Abramowitz70}
M. Abramowitz and I. A. Stegun, \emph{Handbook of Mathermatical
Functions} (Dover, New York, 9th printing, 1970).


\bibitem{Parr94}
R. G. Parr and W. Yang, \emph{Density-Functional Theory of Atoms and
Molecules}, (Oxford University Press, 2nd ed.\ 1994).

\bibitem{Morton05}
K. Morton and D. Mayers, \emph{Numerical Solution of Partial
Differential Equations, An Introduction}, (Cambridge University
Press, 2nd ed.\ 2005).

\bibitem{Bloch08}
I. Bloch, J. Dalibard, and W. Zwerger, Rev.\ Mod.\ Phys.\ {\bf 80},
885 (2008).



\bibitem{McMahon12}
J. M. McMahon, M. A. Morales, C. Pierleoni, and D. M. Ceperley,
Rev.\ Mod.\ Phys.\ {\bf 84}, 1607 (2012).

\bibitem{Pollet12}
L. Pollet, Rep.\ Prog.\ Phys.\ {\bf 75}, 094501 (2012).


\bibitem{Sciver12}
S. W. van Sciver, \emph{Helium Cryogenics} (Springer, New York, 2nd
ed., 2012).

\bibitem{Attard97a}
P. Attard, O. G. Jepps, and S. Mar\v{c}elja,
``Information Content of Signals using
Correlation Function Expansions of the  Entropy'',
Phys.\ Rev.\ E {\bf 56}, 4052--4067 (1997).


\bibitem{Attard98}
P. Attard,
``Information Entropy due to Multi-level Digitisation'',
Mol.\ Phys.\ {\bf 95}, 439--447 (1998)


\bibitem{Attard99}
P. Attard,
``Markov Superposition Expansion for the Entropy
and Correlation Functions in Two and Three Dimensions'',
in `Statistical Physics on the Eve of the Twenty-First Century',
M. T. Batchelor and L. T. Wille (eds),
(World Scientific, Singapore, 1999).

} 

\end{thebibliography}
\end{document}